# Enjeux citoyens de la formation à l'intelligence artificielle

Margarida ROMERO, Universitat Internacional de Catalunya

https://orcid.org/0000-0003-3356-8121 ; margarida.romero@gmail.com

**Résumé**

Dans ce chapitre nous abordons les enjeux citoyens de l'IA en éducation, notamment en ce qui concerne les élèves, les enseignant.e.s et tout autre acteur éducatif face à l'intégration de l'IA. Nous présentons d'abord la manière d'acculturer et former à l'IA, ainsi que et les différentes stratégies pour développer une réflexion sociocritique en matière de formation à l'IA, afin de déterminer quels sont les usages pertinents et éthiques à privilégier. Pour ce faire, dans un deuxième temps, nous évoquons des compétences de pensée critique et de pensée informatique susceptibles d'être mobilisées dans le cadre de certaines activités éducatives soutenues par l'IA, selon le degré d'engagement créatif et transformatif que ces activités raquèrent.

Mots-clés : intelligence artificielle (IA), enseignement, compétences, engagement.

**Introduction**

Jusqu'à l'année 2022, l'intelligence artificielle (IA) était souvent perçue comme une technologie énigmatique, réservée aux initiés et éloignée tant du grand public que des enceintes éducatives. Cependant, au fil des années récentes, les applications reposant sur les algorithmes d'IA ont connu une expansion spectaculaire (Chiu et al., 2023). Les domaines liés à l'IA se sont immiscés dans une multitude de secteurs et d'industries, incluant la santé, la finance, la sécurité et la gouvernance, et dans une certaine mesure, l'éducation. Si certains domaines comme la finance et la sécurité ont entrepris d'exploiter activement le potentiel de l'IA pour améliorer l'efficacité, la précision et la célérité de leurs opérations dans une logique d'optimisation de processus (Fares et al., 2022), l'éducation demeure un secteur très différent quant au type de processus qui peuvent être soutenus par l'IA (Borenstein & Howard, 2021). En effet, l'éducation ne saurait être comparée à d'autres domaines d'application de l'IA puisqu'elle requiert une attention tout particulière dans la manière dont il est possible d'aborder de considérer les aspects de l'IA qui peuvent apporter une valeur ajoutée aux processus d'enseignement, d'apprentissage et de relation famille-école. Ainsi, le choix des applications de l'IA dans le cadre éducatif suscite une réflexion qui transcende la seule dimension technologique. Il s'agit avant tout d'une réflexion qui touche à la culture, à la citoyenneté et aux valeurs sociétales (Robinson, 2020) et éducatives (Adams et al., 2021) qu'il importe de préserver et de soutenir au travers des diverses activités d'enseignement et d'apprentissage soutenues par l'IA.

Face au défi que représente l'acculturation et la formation à l'IA pour l'ensemble des acteurs éducatifs, ce chapitre s'amorce en explorant les enjeux relatifs à l'acculturation et au développement des compétences en IA. Par la suite, nous aborderons les enjeux du développement des compétences de de pensée critique et pensée informatique pour soutenir une intégration de l'IA en éducation à visée citoyenne.

La figure 1 synthétise les sujets abordés dans ce chapitre, allant des opportunités aux enjeux critiques de l'IA en éducation, tout en mettant l'accent sur les compétences de pensée critique et pensée informatique pour permettre le développement de différentes activités d'apprentissage soutenues par l'IA. Les compétences #5c21 comprennent la pensée critique, la pensée informatique, la résolution de problèmes, la créativité et la collaboration.

Figure 1

*De l'aliénation citoyenne face à l'automatisation et les prises de décision assistées par l'IA à la collaboration humain-IA à visée transformative et citoyenne*

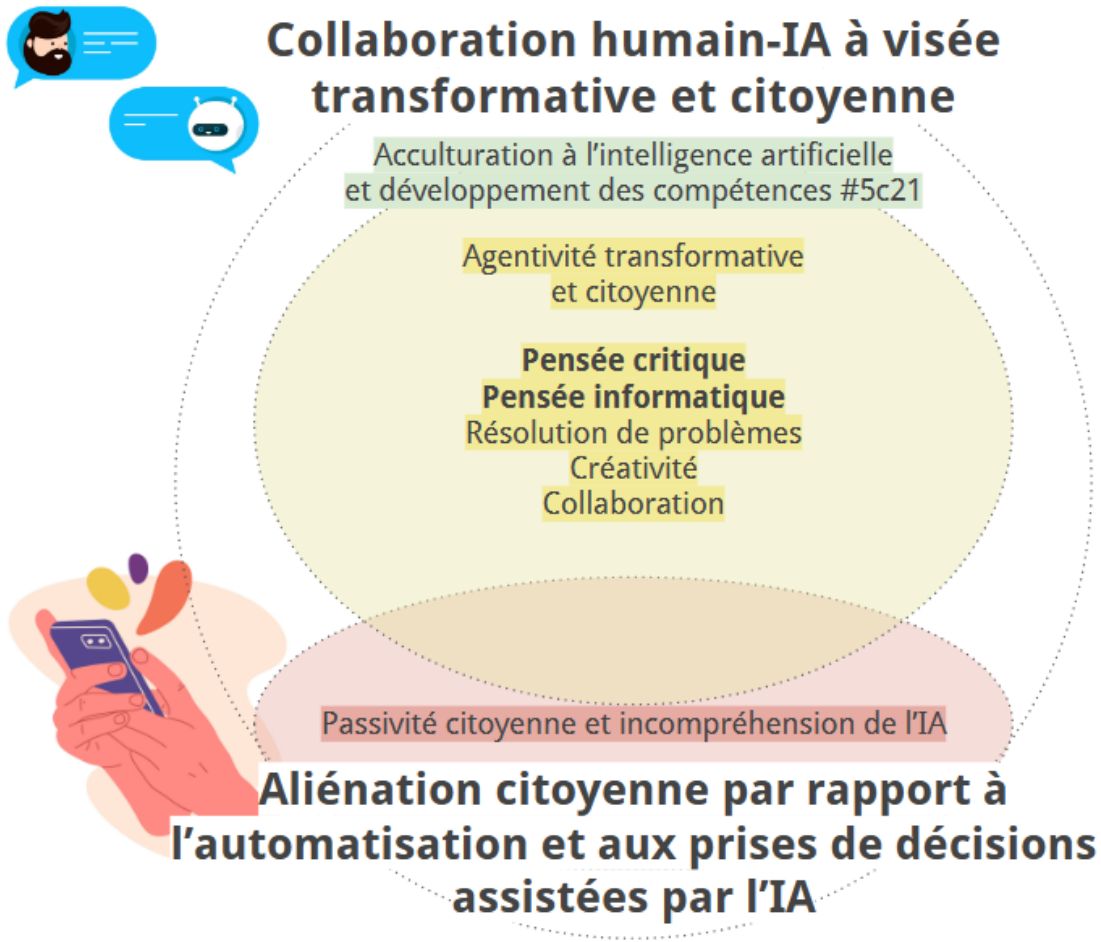

## 1. Acculturation à l'IA

La démocratisation de technologies IA génératives, telles que *ChatGPT* ou *MidJourney*, a ouvert la porte à des questionnements profonds de la part du grand public, des acteurs éducatifs et de la recherche en éducation (Chiu et al., 2023). Ces questionnements englobent non seulement les implications de l'IA dans le cadre professionnel et personnel, mais aussi les défis qu'elles suscitent au sein des activités éducatives, citoyennes et démocratiques (Coeckelbergh, 2022; Holmes et al., 2022). Face à cette évolution, une nécessité impérieuse se dessine : celle de sensibiliser et d'acculturer l'ensemble des citoyens à l'IA afin de leur permettre non seulement d'avoir une approche socio-critique en éducation, mais aussi de développer des usages leur permettant de soutenir des activités technocréatives et même transformatives (Alexandre, Comte, Lagarrigue, & Viéville, 2023; Romero et al., 2023). Une telle sensibilisation et acculturation à l'IA vise à offrir à chacun des citoyens les connaissances et les compétences nécessaires pour appréhender les utilisations actuelles de l'IA dans leur quotidien, mais aussi de permettre d'envisager la manière dont ils pourraient s'en servir en bénéfice de leurs activités d'apprentissage, professionnelles ou citoyennes. Comme indique la devise de la *Maison de l'Intelligence Artificielle* (MIA), l'IA n'est pas magique, mais c'est bien un enjeu à la fois technologique que citoyen et sociétal auquel il faut s'acculturer (Alexandre, Comte, Lagarrigue, & Viéville, 2023; Romero et al., 2023).

**2. Formation à l'IA des élèves et des enseignants**

Dans une perspective de formation initiale et continue des enseignants du primaire, du secondaire et de l'éducation supérieure, nous considérons que la compréhension des mécanismes de base en lien à l'intelligence artificielle représente un enjeu majeur de la profession enseignante (Durampart et al., 2023; Romero et al., 2023) et plus largement, de celle de tout citoyen. En effet, en comprenant les concepts de base de l'IA (Alexandre, Comte, Lagarrigue, Mercier, et al., 2023), les citoyens peuvent mieux comprendre comment ces technologies fonctionnent et comment elles peuvent affecter leur vie. Par exemple, en comprenant comment les algorithmes d'IA prennent des décisions, les citoyens peuvent mieux comprendre comment les entreprises et les gouvernements utilisent les données pour prendre des décisions qui les affectent. De plus, en étant plus conscients des avantages et des limites de l'IA, ils peuvent contribuer à façonner le développement de ces technologies de manière plus responsable et éthique. Cela peut aider à éviter les biais et les discriminations dans l'IA (Cheuk, 2021; Fountain, 2022), ainsi qu'à protéger la vie privée et la sécurité des données des utilisateurs (Woolf et al., 2013). S'acculturer à l'IA peut aider les citoyens à devenir pas uniquement des consommateurs de technologies mais à être des co-créateurs avec l'intelligence artificielle pour permettre de développer une collaboration IA-citoyen qui soit au bénéfice des objectifs éducatifs et du bien commun.

Parmi les stratégies qui peuvent être mises en place, nous pourrons considérer les interventions formatives (Engeström, 2011) réalisées auprès des élèves de l'enseignement secondaire dans le but de comprendre leurs représentations en lien à l'IA à la suite de deux activités formatives. L'une des activités, développées dans le cadre du GTnum Scol_IA[1] "Renouvellement des pratiques numériques et usages créatifs du numérique et IA", soutenu par la Direction du Numérique Éducatif (DNE) du Ministère de l'Éducation en France, mène à la compréhension du concept d'algorithme. Ainsi, par l'entremise d'une activité combinant l'analyse narrative et l'écriture d'une code informatique, les élèves ont pu comprendre le concept d'algorithme, d'automate, de robot et de programmation, en réalisant un projet de théâtre interactif à partir du conte *Vibot le Robot* (Romero et al., 2016; Romero & Chiaruttini, 2022). La deuxième activité développée dans le GTnum Scol_IA est basée sur le *Jeu de Nim* et permet de comprendre le fonctionnement d'un algorithme de tri à partir d'une activité d'apprentissage de l'informatique débranchée. À partir de l'analyse des représentations des élèves avant et après ces deux activités nous avons pu observer comment les élèves démystifient les capacités des systèmes artificiels et développent une meilleure compréhension des spécificités de l'intelligence humaine en relation à celle de l'IA (Heiser et al., 2021; Romero, 2020).

---

[1] https://scoliablog.wordpress.com/

## 3. Enjeux de pensée critique pour l'intégration de l'IA en éducation

Les enseignants sont au cœur de la qualité éducative et de la relation avec les apprenants et, dans des étapes éducatives initiales, également de leurs familles. Dans l'activité des enseignants, il est essentiel de considérer et de soutenir leur capacité à prendre des décisions sur les technologies qui peuvent leur être utiles, ou pas. Pour cela, la clé est la formation initiale et continue des enseignants (Salas-Pilco et al., 2022). Les systèmes éducatifs qui investissent davantage dans la formation des enseignants ont tendance à mieux performer dans les évaluations internationales (Tonga et al., 2022) et permettent aux enseignants de développer un niveau d'autonomie plus important. Au contraire, le manque de formation est une difficulté pour le développement des compétences des enseignants et leur capacité à développer leur manière de manière autonome et éclairée par la recherche. D'autre part, il ne s'agit pas uniquement de former à des compétences procédurales d'usages de certaines technologies, mais de former à la culture et les compétences numérique qui puissent permettre aux enseignants de prendre des décisions éclairées face aux choix d'intégration des activités d'apprentissage et des technologies pour les soutenir. Face aux opportunités mais aussi aux risques entraînées par les technologies dites d'IA, les enseignants doivent pouvoir disposer d'une acculturation à l'IA et développer la pensée critique face à ce type de technologies et leur rôle en éducation pour pouvoir décider ce qui est le plus pertinent dans leurs classes. L'étude basé sur la capacité des enseignants à faire des choix d'intégration de l'IA basé sur l'adaptation du modèle TPACK (Celik, 2023) a permis observer que les connaissances sur l'IA et sur l'enseignement de l'IA sont importantes pour que les enseignantes intègrent l'IA de manière pertinente en contexte éducatif.

La pensée critique est un élément caractéristique de l'intelligence humaine que nous ne retrouvons pas dans les intelligences dites artificielles. Ennis (1989) définit la pensée critique comme « une pensée réflexive raisonnable axée sur la décision de savoir quoi croire ou faire » (p.10) ce qui nous amène à considérer que cette pensée réflexive doit précéder une décision sur notre jugement de la qualité d'une information (ou désinformation) mais aussi la manière d'agir face à des décisions comme celles qui peuvent concerner, pour un enseignant, l'intégration, ou pas, d'une nouvelle technologie en classe. Dans la définition d'Ennis il est considéré de manière implicite le fait que cette pensée réflexive est réalisée par un humain, mais nous souhaitons préciser davantage des aspects implicites de cette définition en considérant le caractère autonome de la réflexion liée à la pensée critique, mais aussi au système de valeurs et de jugements propres de l'individu. Ainsi, nous définissons la pensée critique comme la capacité à développer une réflexion autonome, permettant l'analyse d'idées, de connaissances et de processus en lien avec un système de valeurs et de jugements propres (Romero, Lille, et al., 2017). C'est une pensée responsable, fondée sur des normes et sensible aux circonstances et aux opinions des autres. Avec cette considération, l'enseignant doit faire preuve d'une pensée autonome, qui est formée non pas uniquement par les avis majoritaires, mais par une capacité de prendre de la perspective en lien au système de valeurs propre. Ainsi, l'enseignant peut considérer, que malgré une certaine "mode" technologique, il décide de ne pas intégrer un certain usage technologique car celui-ci n'apporte pas assez de valeur ajoutée à l'activité d'enseignement ou d'apprentissage. La capacité à développer également une compréhension du propre système de valeurs éducatives et citoyennes, et pouvoir le mettre en perspective avec les potentiels enjeux éthiques mais aussi d'intérêts commerciaux liés au numérique éducatif, est également pour le développement de la pensée critique de l'enseignant face aux choix d'intégration du numérique, en général, et des technologies d'intelligence artificielles, en particulier.

La pensée critique est absente des systèmes artificiels du fait du manque de sens commun et de sensibilité interpersonnelle (Romero, 2023). La pensée critique qui tente d'être développée dans les systèmes artificiels est un simulacre basé sur la modélisation, toujours approximative, des éléments permettant de simuler cette capacité intrinsèquement humaine (Alexandre et al., 2020). Lorsque nous examinons les résultats de l'apprentissage machine, nous pouvons être effarés par les résultats politiquement incorrects qu'ils ont pu produire en réponse à des images ou des réponses textuelles pouvant être qualifiées de discriminatoires. Dans un système d'apprentissage machine, les résultats peuvent introduire des éléments racistes et sexistes en «apprenant» des données humaines, du fait du manque de valeurs propres et de la sensibilité intersubjective des systèmes artificiels. Dans ce cas, cette soi-disant IA n'est pas responsable et exacerbe certaines dérives que l'humain peut produire (Cheuk, 2021; Fountain, 2022), tout en les limitant et les corrigeant grâce à ses critères et à sa sensibilité aux autres. Ces résultats n'ont pas de valeur morale pour le système artificiel et des humains doivent venir corriger la valeur morale des résultats avant que le système puisse être proposé au grand public (Tubaro & Casilli, 2019).

Le rapport de Villani et al. (2018) évoque la pensée critique face à ces technologies, tant sur les aspects éthiques que sur le besoin de développer «l'esprit critique» dans l'enseignement de ces sujets. D'autre part, le rapport met l'accent sur l'importance de la créativité dans l'éducation comme un moyen de préparer les citoyens aux défis que ces algorithmes rendent possibles. L'éducation basée sur le numérique, en particulier à travers des méthodes critiques, créatives et participatives, peut également aider à développer un rapport à l'informatique qui permet aux citoyens de démystifier l'IA, de développer une exigence éthique et d'adopter une attitude éclairée (acceptation ou non de ce qui sera utilisé dans leurs activités personnelles, sociales ou professionnelles). Soutenir la pensée critique des enseignants est dans cette ère de l'IA un enjeu critique et essentiel afin qu'ils puissent, à leur tour, soutenir celles de leurs élèves (Romero et al., 2023). Mais les exigences ne s'arrêtent pas à la pensée critique, il est également essentiel que les enseignants développent la pensée informatique pour pouvoir mieux apprendre l'IA, ce que nous aborderons dans la prochaine section de ce chapitre, et la capacité à concevoir et orchestrer des activités de collaboration humain-IA à une visée transformative, comme nous aborderons dans la dernière section de ce chapitre.

**4. Développement de la pensée informatique pour l'appréhension de l'IA**

Outre la pensée critique, pour comprendre et pouvoir faire usage de l'IA pour des processus d'enseignement et d'apprentissage de manière pertinente, il est nécessaire d'avoir un minimum de compréhension des fondements de l'informatique et la manière d'utiliser ces fondements pour résoudre des problèmes. Ceci est possible par le développement de la pensée informatique. En 2006, Wing définit la «pensée informatique» (*computational thinking*) comme la capacité à utiliser des processus informatiques pour résoudre des problèmes dans n'importe quel domaine. Pour son développement, les apprenants (dès la maternelle et à tous les âges) peuvent combiner l'apprentissage des concepts et des processus informatiques relevant de la «littératie numérique» (objets, attributs, méthodes, patrons de conception, etc.) avec une démarche de résolution de problèmes créative faisant appel aux concepts et aux processus informatiques (Romero, Lepage, et al., 2017). La pensée informatique engage une démarche de résolution de problèmes et un raisonnement systématique et structuré, essentiel à la compréhension des concepts fondamentaux de l'informatique et de l'IA. Elle repose sur des compétences telles que l'analyse, la décomposition de problèmes, la reconnaissance de modèles, la conception algorithmique, et la pensée algorithmique. Des initiatives telles que Class'Code en France ou CoCreaTic au Québec ont élaboré des ressources et créé des communautés pour soutenir cette approche, où l'objectif n'est pas d'apprendre le «codage» pas à pas, mais de résoudre de manière créative et contextuellement sensible des problèmes. En d'autres termes, dépasser le simple codage permet de s'ancrer dans une démarche plus vaste de programmation créative. Celle-ci engage les apprenants dans un processus critique et créatif de résolution de problèmes faisant appel aux concepts et aux processus informatiques. Il ne s'agit pas de décoder pour coder, ou d'écrire des lignes de code les unes après les autres, mais de développer une approche de résolution de problèmes complexes qui implique une analyse réflexive et empathique de la situation, de sa représentation et de l'opérationnalisation d'une solution profitant des stratégies métacognitives liées à la pensée informatique. La pensée informatique, telle que présentée par Wing (2009, para. 3), se révèle comme un ensemble d'attitudes et de connaissances universellement applicables, au-delà de l'usage des machines. Elle englobe, par exemple, la capacité à nommer de manière pertinente les objets et à expliciter leur type ou catégorie pour les manipuler correctement, à maîtriser la complexité d'un grand problème ou d'un système en le hiérarchisant, à pouvoir spécifier dans ses moindres détails un procédé pour qu'il puisse s'exécuter sans ambiguïté de manière mécanique, etc. Le terme de pensée informatique est utilisé pour indiquer que l'on souhaite non seulement initier à la programmation, mais permettre aux jeunes de prendre du recul par rapport au numérique et positionner l'apprentissage de l'informatique comme une compétence transdisciplinaire s'appuyant sur le numérique pour développer des stratégies de pensée. La pensée informatique se veut ainsi émancipatrice, visant à former des citoyennes et citoyens éclairés, capables d'esprit critique mais également dotés d'une capacité de compréhension et de (co)création avec la «panoplie d'outils numériques» évoquée par Wing au regard de la pensée informatique. Le développement d'une approche critique et créative du numérique en général, et de l'IA en particulier, par le biais de la pensée informatique permet aux apprenants d'aller au-delà d'une posture d'utilisateurs qui pourrait percevoir l'IA comme une boîte noire pleine de mystères, de dangers ou d'espoirs illimités. Comprendre les enjeux de l'analyse de problèmes liés à des situations problématiques ancrées dans des contextes socio-culturels précis (par exemple, les enjeux migratoires) est une manière de voir l'informatique à la fois comme une science et une technologie qui, à partir des limites et des contraintes de nos modélisations d'un problème, tente de fournir des réponses nourries de données de plus en plus massives, sans pour autant pouvoir être considérées comme pertinentes ou valides sans l'engagement du jugement humain. Ainsi, la pensée informatique nous permet d'appréhender les fondements de l'IA en nous aidant à comprendre comment les données sont traitées et interprétées par les algorithmes (Romero, 2018). Elle nous permet d'appréhender la logique derrière ces algorithmes, les transformations des données, les apprentissages, et les prises de décision. Sans cette compréhension, il est difficile d'appréhender pleinement le fonctionnement de l'IA. En lien à l'agentivité transformative visée par l'éducation à la citoyenneté (Barton, 2012), la pensée informatique nous offre également la capacité de concevoir des solutions innovantes et efficaces aux problèmes complexes. D'autre part, le développement de la pensée informatique est un élément nécessaire pour permettre de faire preuve d'esprit critique et éthique dans le développement et l'utilisation de l'IA. Elle nous permet de réfléchir de manière approfondie aux conséquences sociales, économiques et éthiques de l'IA par la compréhension des processus et concepts clés de l'informatique. En comprenant les algorithmes et les biais potentiels, nous pouvons concevoir des systèmes plus équitables et justes, en minimisant les risques pour notre société, tant d'un point de vue de la prise de décision assistée par l'IA (Holmes et al., 2022; Vereschak et al., 2021) que des enjeux démocratiques liés à la génération de contenus et d'informations assistés par l'IA (Moran & Shaikh, 2022). La pensée informatique est également essentielle pour la compréhension et l'intégration réussie de l'IA dans notre quotidien. Elle forme la base pour développer des compétences cruciales dans un monde de plus en plus numérique et automatisé. En investissant dans l'apprentissage et la promotion de la pensée informatique, nous nous préparons à un avenir où l'IA peut être utilisée de manière responsable et bénéfique pour tous, ce qui s'inscrit dans des démarches d'usage de l'IA pour le bien commun (#AI4good) et le développement d'objectifs de développement durable (augMENTOR, 2023; Sætra, 2021).

**5. Du citoyen passif face à l'IA à l'agentivité citoyenne soutenue par l'IA**

Dans l'évolution des rapports entre les citoyens et l'IA, il est important de considérer les pédagogies créatifs (Lin, 2011) dans le but de développer un rapport citoyen-IA qui soit basé sur une agentivité transformative et citoyenne où chaque individu puisse agir pour le bien commun et sociétale, mais aussi pour son développement personnel et professionnel en tirant le meilleur parti de l'IA et pas simplement en subissant les avancées de l'IA de manière passive comme un consommateur. Dans ce contexte, il s'agit de considérer la manière dont il peut s'opérer le passage fondamental d'apprenant et citoyen passif face à l'IA bénéficiant de services et d'informations générées par l'IA, à l'agentivité citoyenne soutenue par l'IA, à celui d'acteur des usages de l'IA dans une visée citoyenne. Cette transition souligne l'importance de faire évoluer les usages dits passifs vers des usages cocréatifs de l'IA. Dans ce but, nous avons développé le modèle #PPai6 (Romero, 2023; Septiani et al., 2023), qui présente des activités d'apprentissage en lien à l'IA où il est possible distinguer six niveaux différents d'usages créatifs de l'IA, comme présentés dans la figure 2.

Figure 2

*De l'usage passif des contenus générés par l'IA à la créativité transformative et l'apprentissage expansif soutenu par l'IA*

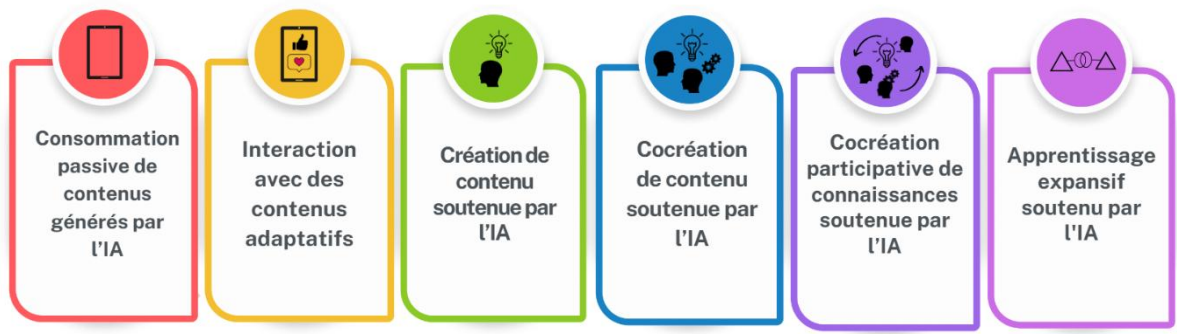

Romero © 2023

Le modèle Passif-Participatif des usages de l'IA en éducation #PPai6 présente six niveaux d'engagement créatif dans l'éducation citoyenne à l'IA. Pour exploiter pleinement les avantages de l'IA avec une visée citoyenne et de créativité transformative (Lim et al., 2023; Romero, 2019), le modèle #PPai6 est articulé en six niveaux distincts d'engagement créatif.

- *Consommation passive de contenus générés par l'IA*. Au premier niveau, l'apprenant est un consommateur passif de contenu généré par l'IA, ne comprenant pas forcément son fonctionnement.
- *Interaction avec des contenus adaptatifs*. Au deuxième niveau, l'interaction entre l'apprenant et le contenu généré par l'IA est encouragée, permettant à l'IA de s'ajuster en fonction des actions de l'apprenant.
- *Création de contenu soutenue par l'IA*. Le troisième niveau implique la création individuelle de nouveau contenu par l'apprenant en utilisant des outils d'IA.
- *Cocréation de contenu soutenue par l'IA*. Au quatrième niveau, la création collaborative de contenu est mise en avant, où une équipe d'apprenants utilise des outils d'IA pour produire du nouveau contenu de manière collective.

- *Cocréation participative de connaissances soutenue par l'IA*. Au cinquième niveau, l'accent est mis sur la co-création participative des connaissances, où une équipe utilise des outils d'IA en collaboration avec divers acteurs pour aborder des problèmes complexes.
- *Apprentissage expansif soutenu par l'IA*. Enfin, au sixième niveau, l'IA est un catalyseur pour un apprentissage expansif, permettant aux participants de résoudre des problèmes complexes et de transformer les situations problématiques grâce à des outils d'IA qui facilitent la visualisation, la modélisation et l'action.

En comprenant ces niveaux d'engagement, les enseignants peuvent choisir des stratégies pédagogiques adaptées pour encourager un apprentissage plus actif, collaboratif et créatif. C'est un modèle qui s'inscrit dans une démarche de développement des compétences transversale de pensée critique et de pensée informatique, dans une visée citoyenne d'éviter subir les impacts de l'IA, mais de permettre aux apprenants et citoyens de tout âge de devenir des acteurs impliqués, voire co-créateurs des usages de l'IA, dans son développement, son déploiement et sa réglementation éthique dans les différents domaines de la vie. Il incite à repenser les interactions citoyennes avec les technologies et à encourager une participation active dans la conception et l'orientation des systèmes d'IA. Ce passage vers des usages cocréatifs s'inscrit dans une logique d'engagement proactif, de collaboration et de co-construction entre les citoyens des usages et régulations de l'IA. Les enseignants ont un rôle clé pour permettre de modeler des usages de l'IA qui ne sont pas juste des aides cognitives ou d'autonomisation de certains tâches, mais de permettre d'appréhender l'utilisation de l'IA comme un outil au service de l'agentivité transformative et de la mise en lumière des enjeux éthiques et citoyens des usages du numérique.

La figure 3 présente le modèle #PPai6 dans cette visée citoyenne, qui peut favoriser tant l'aliénation face à la délégation aux IA de certaines prises de décision (Vereschak et al., 2021) que le soutien à l'agentivité transformative des citoyens.

Figure 3.

*De l'aliénation citoyenne à la créativité transformatrice et l'apprentissage expansif soutenu par l'IA au service de la société.*

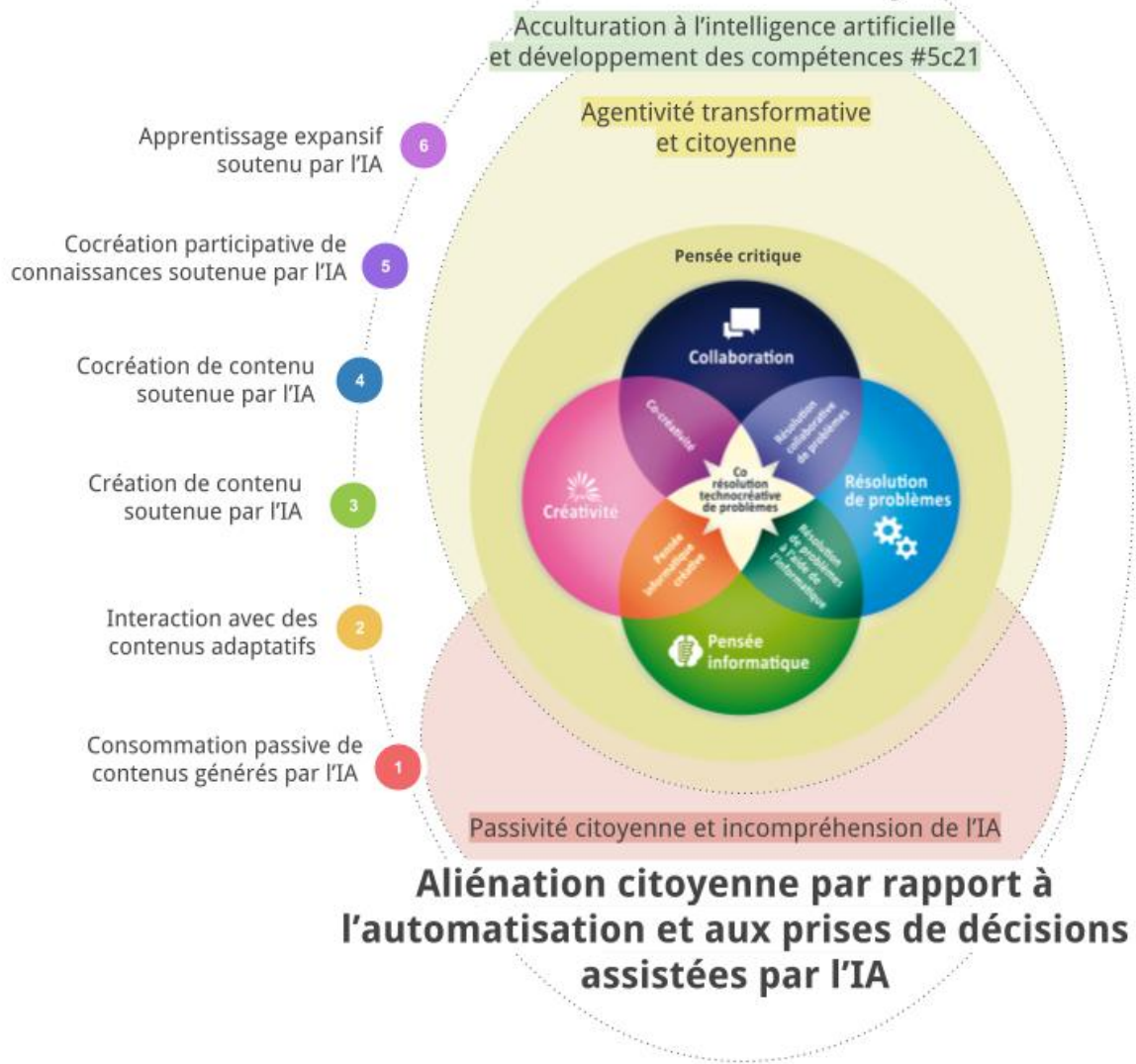

Pour que l'intégration de l'IA en éducation et dans la société puisse se faire sous une approche de créativité transformative (Alexandre, Comte, Lagarrigue, & Viéville, 2023; Romero et al., 2023) et d'apprentissage expansif il faut développer, tout d'abord, réaliser des actions d'acculturation et formation à l'IA tnat des enseignants que des élèves, pour ensuite sensibiliser aux enjeux éthiques, aux capacités et aux limites de l'IA dans les différentes domaines d'activité citoyenne, académique, personnelle, et professionnelle. Cette éducation doit permettre aux citoyens d'acquérir les compétences nécessaires pour comprendre et contribuer à la conception des usages et la régulation de l'IA, en s'assurant que les valeurs et les besoins de la société soient maintenus en accord avec des principes humanistes et démocratiques.

En guise de conclusion, malgré tout l'emphase mise dans ce chapitre dans les activités d'enseignement et d'apprentissage, il faut également considérer que les pouvoirs publics et les entreprises technologiques ont un rôle crucial à jouer pour développer des technologies et des réglementations permettant de maintenir l'agentivité citoyenne. Ils doivent favoriser la transparence et la responsabilité des systèmes d'IA, encourager la participation citoyenne dans les processus de décision et d'évaluation, et promouvoir des mécanismes de gouvernance qui garantissent une représentation équitable des différents citoyens au long de leur vie. Nous préconisons la coopération entre acteurs éducatifs, développeurs des technologies IA, chercheurs tant en sciences informatiques qu'humaine et décideurs pour identifier les besoins, les priorités et les applications bénéfiques de l'IA en éducation et dans d'autres domaines citoyens essentiels où il est nécessaire de réfléchir de manière critiques à l'intégration, ou pas, de l'IA selon le type d'activités. Le modèle #PPai6 invite à repenser le rôle du citoyen vis-à-vis de l'IA en passant d'une posture passive à une agentivité citoyenne soutenue par l'IA. Cette transition vers des usages cocréatifs représente un enjeu pour construire un futur où l'IA est un levier d'amélioration de la société, guidée par les valeurs et les besoins de ses citoyens, avec la contribution des citoyens en tant qu'acteur et pas uniquement comme consommateurs.

**Références**